\pdfoutput=1
\documentclass[twocolumn]{aastex61}
\usepackage{multirow}
\usepackage{graphics}
\usepackage{amsmath, amsthm, amssymb}
\usepackage{epsfig}
\usepackage{natbib}

\usepackage{graphicx}

\received{....}
\revised{...}
\accepted{...}
\submitjournal{ApJ}

\accepted{May 10, 2017}
\submitjournal{ApJ}

\shorttitle{The dynamical origin of the Local arm}
\shortauthors{L\'epine et al.}

\begin{document}

\title{The dynamical origin of the Local arm and the Sun's trapped orbit}

\correspondingauthor{Jacques L\'epine}
\email{jacques@astro.iag.usp.br, tatiana.michtchenko@iag.usp.br, douglas.barros@iag.usp.br, rss.vieira@usp.br}

\author{Jacques R.\,D. L\'epine}
\affil{Universidade de S\~ao Paulo, IAG,  \\
Rua do Mat\~ao, 1226, Cidade Universit\'aria, \\
05508-090 S\~ao Paulo, Brazil}

\author{Tatiana A. Michtchenko}
\affil{Universidade de S\~ao Paulo, IAG,  \\
Rua do Mat\~ao, 1226, Cidade Universit\'aria, \\
05508-090 S\~ao Paulo, Brazil}

\author{Douglas A. Barros}
\affil{Universidade de S\~ao Paulo, IAG,  \\
Rua do Mat\~ao, 1226, Cidade Universit\'aria, \\
05508-090 S\~ao Paulo, Brazil}

\author{Ronaldo S. S. Vieira}
\affil{Universidade de S\~ao Paulo, IAG,  \\
Rua do Mat\~ao, 1226, Cidade Universit\'aria, \\
05508-090 S\~ao Paulo, Brazil}











\begin{abstract}
The Local arm of the Milky Way, a short spiral feature near the Sun whose existence is known for decades, was recently observed in detail with different tracers. Many efforts have been dedicated to elaborate plausible hypotheses concerning the origin of the main spiral arms of the Galaxy; however, up to now, no specific mechanism for the origin of the Local arm was proposed.
Here we explain, for the first time, the Local arm as an outcome of the spiral corotation resonance, which traps arm tracers and the Sun inside it.
We show that the majority of maser sources belonging to the Local arm, together with the Sun, evolve inside the corotation resonance, never crossing the main spiral arms but instead oscillating in the region between them. This peculiar behavior of the Sun could have numerous consequences to our understanding of the local kinematics of stars, the Galactic Habitable Zone, and the Solar System evolution.
\end{abstract}



\section{Introduction}
\label{intro}

The Milky Way is considered to be a Grand Design spiral galaxy, as most works on spiral arm tracers indicate \citep{georgelinGeorgelin1976AA, levine2006Science, houHan2009AA, houHan2014AA, reidEtal2014ApJ}. At regions not too far from the Sun, the geometry of the spiral structure is well known. The main features are the extended Sagittarius-Carina arm, which passes at an inner galactic radius (compared to the Sun), and the Perseus arm, at an outer radius. These arms are revealed by molecular clouds, H\,{\scriptsize II} regions, OB stellar associations, and open clusters, among other tracers \citep{georgelinGeorgelin1976AA,houHan2014AA,bobylevBajtkova2014MNRAS,
reidEtal2014ApJ}.
The most accurate picture of the spiral arms in the Galactic plane is given by maser sources associated with star forming regions, whose distances are obtained by recent Very Long Baseline Interferometry (VLBI) parallax measurements \citep{reidEtal2014ApJ}; in this case, the calculation of the distances needs no assumption about the rotation curve and interstellar extinction. Inside the solar circle, the position of the arms with respect to the Sun can be determined by the directions of lines of sight tangent to them \citep{vallee2016AJ}.

Midway between the Sagittarius-Carina and Perseus arms, and close to the Sun, there is a short structure, the Local arm  \citep[or Orion Spur; see][]{xuEtal2013ApJ, bobylevBajtkova2014MNRAS, houHan2014AA, reidEtal2014ApJ, xuEtal2016Science}, which is also precisely traced by the VLBI maser sources and other tracers \citep{houHan2014AA}. Even though this arm is known for decades \citep{morganEtal1952AJ, morganEtal1953ApJ, bokEtal1970IAUS, georgelinGeorgelin1976AA} and extensively studied by observational means, there was no explanation for its origin up to now.

The main arms are usually interpreted as the crowding of successive stellar orbits of different radii that produces increased stellar densities and creates elongated valleys of gravitational potential \citep{kalnajs1973PASAu,contopoulosGrosbol1986AA,junqueiraEtal2013AA}.
The self-consistent portrait of the Galaxy shows the spiral pattern rotating with a constant angular velocity (the pattern speed $\Omega_p$), similar to a rigid body.
The pattern speed $\Omega_p$ determines the corotation circle, at which stars and gas rotate around the galactic center with the same average velocity of the spiral arm pattern; the average velocity of the stars is given by the rotation curve, which is relatively flat for the Milky Way \citep{clemens1985ApJ, sofueHonmaOmodaka2009PASJ}. Around the corotation circle, the islands of orbital stability appear in the form of banana-like regions in the effective potential maps; they are associated with the corotation resonance  \citep{contopoulos1973ApJ,michtchenkoVieiraBarrosLepine2017AA}. Hereafter, we will refer to these islands of stability as ``corotation zones''; the number of corotation zones corresponds to the number of spiral arms adopted in the Galaxy model.  Early theoretical studies on the corotation dynamics in disk galaxies predicted a trapped stellar mass inside such corotation zones\footnote{We make clear that our references throughout the paper to ``being inside the corotation resonance/island/zone" actually mean being located in the island of trapped orbits, not lying at a smaller radius with respect to the corotation radius.} \citep{contopoulos1973ApJ,barbanis1976AA}.
Moreover, it was recently found, by means of magnetohydrodynamics simulations, that the gas also remains trapped in the corotation zones \citep{gomezEtal2013MNRAS}.

We have plenty of evidence that we live near the spiral corotation circle
\citep{marochnikEtal1972ApSS,crezeMennessierAA1973,marochnik1983Ap, mishurovZenina1999AA, lepineMishurovDedikov2001ApJ, diasLepine2005ApJ, lepineEtal2011MNRAS}. A direct measurement of the corotation radius was made by \cite{diasLepine2005ApJ} by using a sample of open clusters with known distances, ages and space velocities. Integrating the orbits to the past towards their birthplaces, the authors followed the time displacement of the spiral arms and determined the corotation radius as $R_{CR}=(1.06\pm 0.08)R_0$, where $R_0$ is the galactocentric radius of the Sun. Observations also show that the Sun lies in the vicinity of the Local arm, at a distance smaller than 500\,pc \citep[e.g.][among others]{houHan2014AA}. Owing to the fact that the Sun is close to both the corotation circle and the Local arm, we elaborate a hypothesis on the origin of the Local arm and present it in this paper.

Within the context of a spiral structure with a well-defined corotation radius, we build a Galactic potential model which is composed by an axisymmetric component and a perturbation term due to a four-armed spiral structure. For observationally constrained physical and dynamical parameters, the model
gives rise to four corotation zones. One of these lies between the Sagittarius-Carina and Perseus spiral arms encompassing the position of the Sun; it will be hereafter referred to as the ``local corotation zone''. The closeness to the corotation radius and the superposition of the banana-shaped region over the Local arm position in the Galactic plane allow us to presume a natural connection between observational and dynamical phenomena, namely the Local arm and the local corotation zone. This conjecture is supported by the results of recent gas simulations, which show that the `Local arms' form consistently in the gas density response to the action of an external spiral potential \citep{liEtal2016ApJ}. Moreover, the reported corotation radius of those simulations shows that these arms are relatively close to the corotation circle.

We look for evidence that the mass trapped inside the local corotation zone actually forms the Local arm, and that the Sun probably evolves inside this zone. Using a sample of young objects (maser sources) associated to the Local arm, we study their dynamics by performing numerical integrations of the equations of motion. Our results show that the majority of these objects do not escape from the local corotation zone, indicating that their orbits are trapped inside it. Since these objects are identified as tracers of the Local arm, we can propose that the Local arm is an outcome of the resonant dynamics induced by perturbations due to the main galactic spiral arms on a background axisymmetric disk. The trapping mechanism is similar to the one observed in the Solar System of the Jupiter Trojan asteroids, which are trapped in the L$_4$, L$_5$ Lagrangian solutions for the Sun-Jupiter system \citep[][Chapter 3]{murray1999solar}.
Thus, knowing the mechanism which originates the Local arm, we can elaborate a scenario for its formation and evolution.

To simulate the evolution of the objects inside the Local arm, we adopt a model in which the galactic spiral structure is long-lived, which is likely the case of our Grand-design Milky Way. This model helps us picture the long-term evolution of the solar orbit. A recent dynamical analysis of the neighborhood of the Sun has shown that it may evolve  inside a stable island of the corotation resonance \citep{michtchenkoVieiraBarrosLepine2017AA}. For the observationally constrained galactic parameters of the present paper, the Sun's orbit is found to be trapped in the local corotation zone. In the frame of reference rotating with the spiral pattern, the Sun's orbit evolves oscillating in both radial and azimuthal directions, never crossing the main spiral arms but instead remaining inside the region between them. We discuss the consequences of this orbital behavior of the Sun and stellar objects of the Local arm in terms of habitable zones in the Galaxy and the Solar System evolution.

There is still a debate in the literature between two general lines of thinking concerning the lifetime of spiral arms. Some groups consider that the arms are long-lived, quasi-steady features, and others, based on N-body simulations, consider that the arms are short-living transient structures \citep[e.g.][and references therein]{sellwood2011MNRAS}. Note, however, that there are also N-body simulations that produce long-lived patterns \citep{elmegreenThomassonAA1993, zhang1996ApJ, donghiaEtal2013ApJ}, some of them very recent \citep[e.g.][]{sahaElmegreen2016ApJL}.
Moreover, \cite{fujiiEtal2011ApJ} obtained long-lived patterns in cases when the number of particles employed in N-body simulations was sufficiently large, e.g. $3\times 10^6$ particles. From the observational point of view, \cite{martinez-garciaGonzalez-Lopezlira2013ApJ} analyzed azimuthal age/color gradients across spiral arms for a sample of 13 normal or weakly barred galaxies, and verified that at least $50\%$ of the objects show signatures of long-lived patterns.

It must be emphasized that the existence of the corotation zone needs no assumption about the lifetime of the Galactic spiral structure.
Indeed, once a spiral mode emerges in the Galactic disk, the corotation zones appear instantaneously as a natural consequence of the spiral arms perturbation \citep{contopoulos1973ApJ}. Nevertheless, the knowledge of the timescale is necessary to distinguish between the quasi-steady or transient nature of the main spiral arms; the Local arm structure, with its current features, also depends on this timescale.

The organization of this paper is as follows: In Section~\ref{sec:masers} we describe the sample of objects used to trace the Local arm. In Section~\ref{sec:model} we introduce the model of the Galactic disk and the potential of the spiral arms. In Section~\ref{sec:Hamiltonian} we present the topology of the Hamiltonian and resulting energy levels. We discuss the solar orbit and the evidences for the adopted pattern speed in Section~\ref{sec:patternspeed}, while the analysis of the dynamical map and phase-space structures inside the local corotation zone is done in Section~\ref{sec:map}. In Section~\ref{clusters}, we discuss a second sample of young objects, the Open Clusters.
The discussion in Section~\ref{sec:discussion} includes a scenario for the origin and evolution of the Local arm.

\begin{figure}
\begin{center}
\epsfig{figure=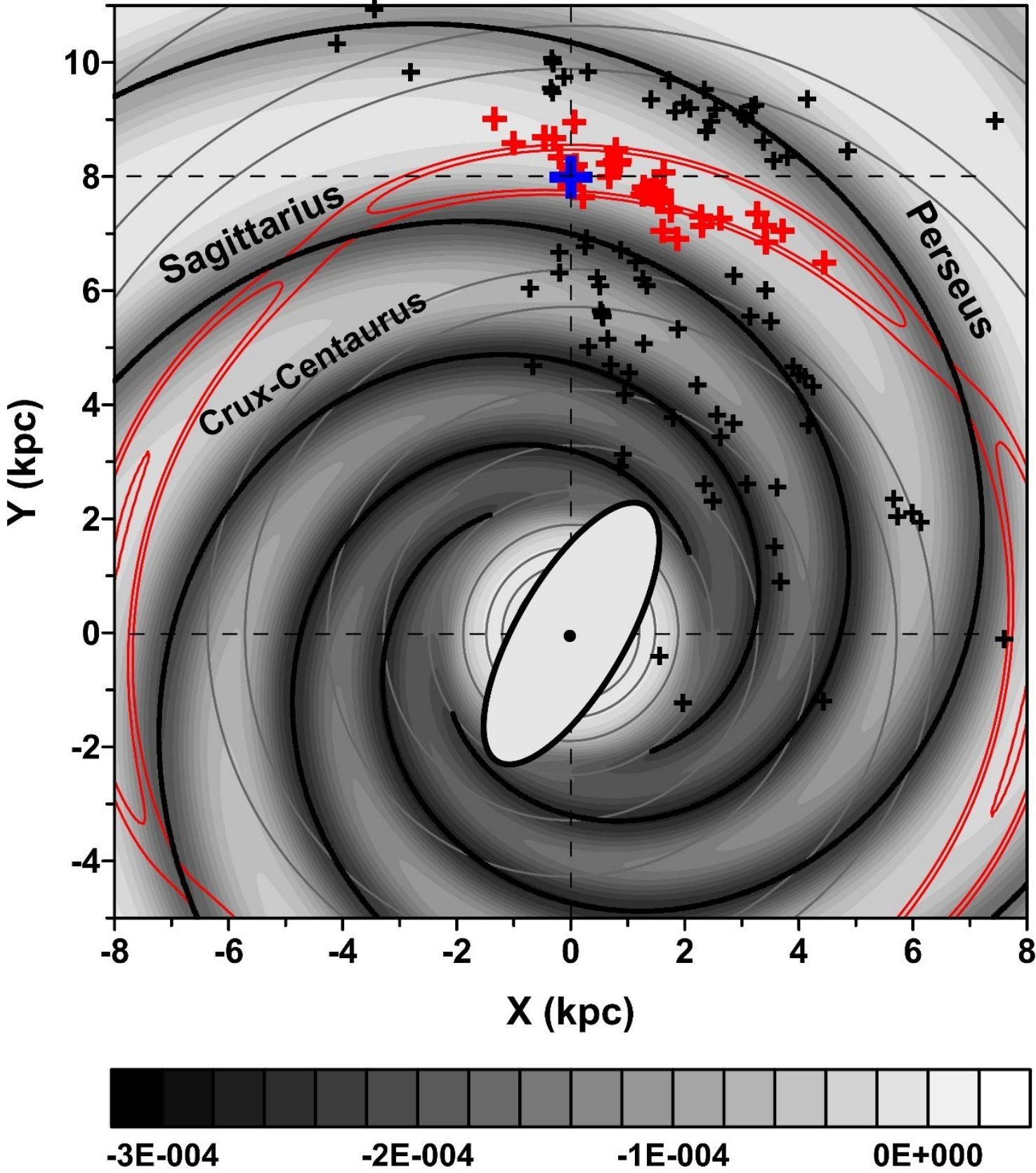,width=0.99\columnwidth ,angle=0}
\caption{Maser sources and Sun in the Galactic $X$--$Y$ plane. The Sun is marked by a blue cross. Red crosses represent masers belonging to the Local arm, whereas black crosses represent masers from the main spiral arms.
The black spiral curves are the loci of the main spiral arms with the physical parameters from Table~\ref{tab:parameters}, except \mbox{$i=+14^\circ$}. 
The distance of the Sun from the Sagittarius arm is 1\,kpc, in the Sun-Galactic center direction. The color bar shows the intensity of the Gaussian shape of the spiral potential, in units of kpc$^2$\,Myr$^{-2}$. The effective potential (Eq.~\ref{eq:effectivepotential}) is represented by the gray levels and the red banana-like levels emphasize the corotation zones. The central bar is only schematic and is not included in our model.
}
\label{fig:bracosXY}
\end{center}
\end{figure}

\section{Maser sources in the Local arm}
\label{sec:masers}

The Local arm tracers are the fundamental tool to analyse the main features of this structure. We consider in this paper data from maser sources associated with regions of star formation as the main tracers of the Local arm. The masers constitute the best set of available data. However, we will also analyse a sample of young Open Clusters in Section~\ref{clusters}.


The present work benefited from the recent publications of trigonometric parallax and proper motion measurements, by means of VLBI techniques, of maser sources associated with High Mass Star Forming Regions (HMSFRs). The observations have been performed by the Bar and Spiral Structure Legacy (BeSSeL) Survey\footnote{http://bessel.vlbi-astrometry.org} key science project, the European VLBI Network\footnote{http://www.evlbi.org}, and the Japanese VLBI Exploration of Radio Astrometry project (VERA)\footnote{http://veraserver.mtk.nao.ac.jp}.
Complementing these data with heliocentric radial velocities from Doppler shifts, we are able to access the full three-dimensional location of each source in the Galaxy as well as their full space velocities relative to the Sun.

\begin{figure}
\begin{center}
\epsfig{figure=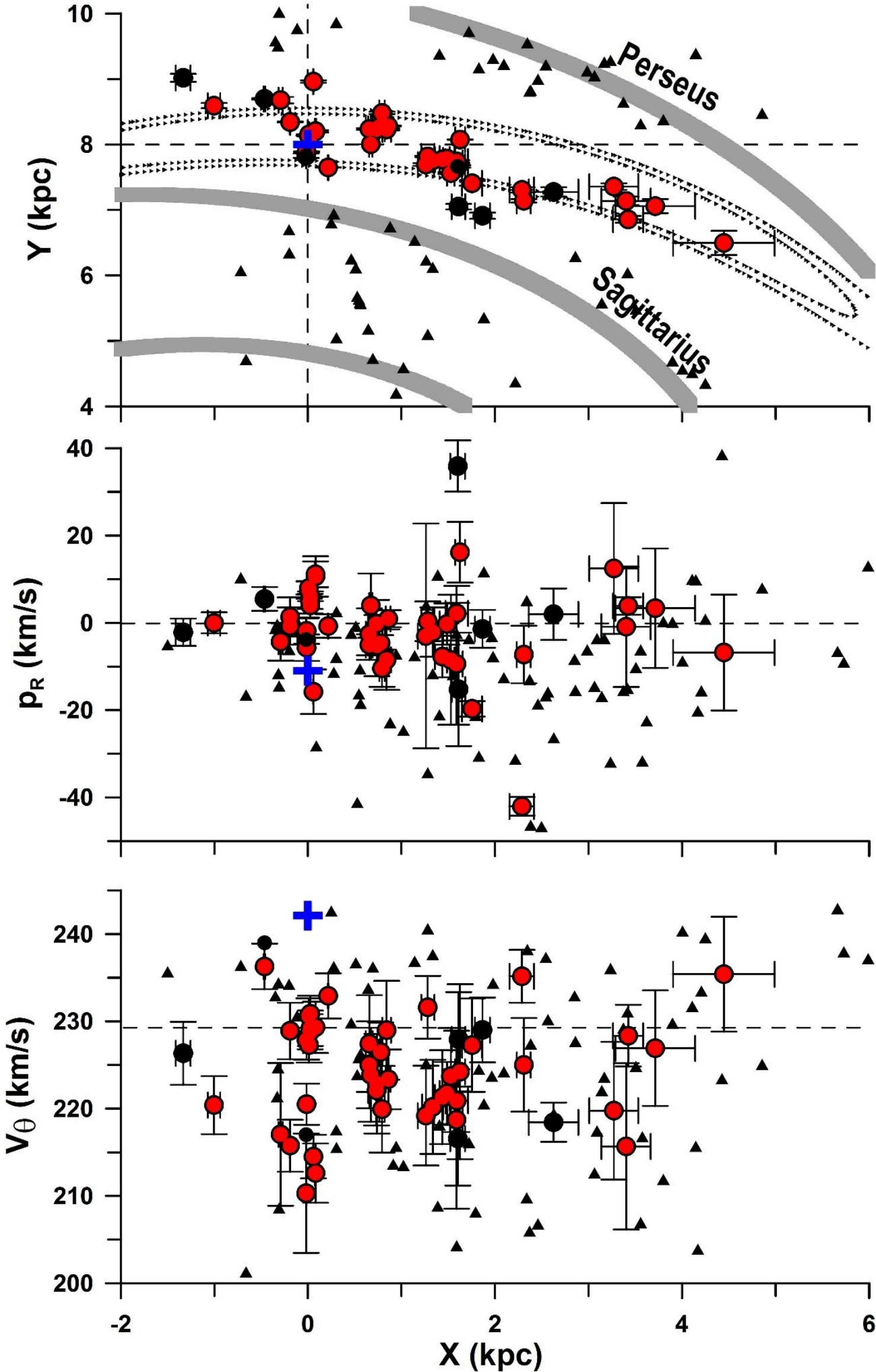,width=0.99\columnwidth ,angle=0}
\caption{ \textit{Top}: $X$--$Y$ current positions of maser sources, with error bars (triangles and filled circles). The filled circles are masers from the Local arm (see Table~\ref{tab:masers} in Appendix): red (black) show masers inside (outside) the corotation resonance. The blue cross represents the Sun's location inside the corotation resonance. The levels of the effective potential (\ref{eq:effectivepotential}) are shown by dotted curves and the loci of the Perseus and Sagittarius arms by thick curves.
\textit{Middle}: Same as the top panel, except for the values of radial momentum $p_R$. The dashed line represents the equilibrium value $p_R=0$.
\textit{Bottom}: Same as the top panel, except for the values of tangential velocity. The dashed line represents the circular velocity of the corotation center, $V_\theta=\Omega_p\,R_{CR}$ (see Section \ref{sec:Hamiltonian}).
}
\label{fig:masersPanels}
\end{center}
\end{figure}

The data for HMSFRs with maser emission, i.e., coordinates, trigonometric parallaxes, proper motion components, LSR radial velocities, and the errors in the measurements,  were obtained from: Table 1 of \cite{reidEtal2014ApJ} (103 sources), and Table 5 of \cite{rastorguevEtal2016} (40 sources). All the tables in the above-cited references contain the original references relative to each individual maser data.
With this sample of maser sources, we can rely on data measured with great accuracy, as it is for both position and velocity determinations.
In addition, since they are associated with massive stars which have short lifetimes, these objects have not moved far away from their birthplaces, so that we can use their positions to trace the spiral arms. The data provide initial conditions for integration of their orbits. In total we have 142 masers\footnote{We noticed that the maser source G109.87+02.11 has entries in both tables of \cite{reidEtal2014ApJ} and of \cite{rastorguevEtal2016}, so we took the data of the latter reference which is from more recent observations.}, of which 47 are situated in the Local arm. The positions of the maser sources in the Galactic plane are displayed in Fig.~\ref{fig:bracosXY} as black and red crosses; the blue cross indicates the position of the Sun. The other features in Fig.~\ref{fig:bracosXY} are explained in the subsequent sections. The choice of which source belongs to the Local arm is the following: from the list of \cite{reidEtal2014ApJ}, we just took the masers identified by these authors as members of the Local arm (24 objects); from the list of \cite{rastorguevEtal2016}, we selected the sources with the lowest distances (on the $X$--$Y$ plane of Fig.~\ref{fig:bracosXY}) to the group formed by the masers of the Local arm from the sample of \cite{reidEtal2014ApJ} (23 objects). The list of \cite{rastorguevEtal2016} already contains some of the masers of the Local arm from \cite{xuEtal2013ApJ} (which are not in the list of \citealt{reidEtal2014ApJ}), and masers from \cite{xuEtal2016Science}.

All the 47 masers of the Local arm are represented by red crosses in Fig.~\ref{fig:bracosXY}, and a list of data for their orbital elements is presented in Table~\ref{tab:masers} in Appendix. The top panel of Fig.~\ref{fig:masersPanels} shows the $X$--$Y$ positions (with error bars) of the masers of the Local arm (red and black circles), and of masers associated with main spiral arms (triangles). 
In the following, we describe the method used for calculating the orbital elements of the masers of the Local arm.
Although the question about whether the velocity of the maser sources are representative of the star forming regions is a matter of debate, we assume, as a working hypothesis, that this is the case here. Therefore, we use the positions and velocities of these sources as initial conditions for the integrations of the orbits of stars belonging to these regions. It should be noted that an important fraction of the masers originate in the circumstellar envelopes of massive stars \citep{inayoshiEtal2013ApJL}, and on the other hand, quite often the spectra of the maser sources present a more intense peak, which is considered to give the ``systemic velocity'' of the compact source, and less intense features that may originate in outflows \citep{ellingsen2004IAUS}. The systemic velocity is then usually considered to be the velocity of the source.

We converted the tabulated LSR radial velocities to heliocentric radial velocities by adding back the components of the standard solar motion \citep{reidEtal2009ApJ}. With the coordinates, parallaxes, proper motions, and heliocentric radial velocities, we calculated the heliocentric $U$ and $V$ velocities for each source, following the formalism described in the Appendix of \cite{reidEtal2009ApJ}. Correcting for the solar peculiar motion from \cite{schonrichBinneyDehnen2010MNRAS} and for the LSR circular velocity $V_0$ (230\,km\, s$^{-1}$, see Section \ref{sec:rotcur}), we calculated the Galactocentric components $V_{\theta}$ and $V_R$ of the space velocity  of each maser source, $V_{\theta}$ being positive in the direction of Galactic rotation and $V_R$ positive towards the Galactic anti-center direction. The uncertainties on $V_{\theta}$ and $V_R$ were obtained by propagation from the uncertainties on the parallaxes, proper motions, heliocentric radial velocities, and the uncertainties of the solar motion.
The middle and bottom panels of Fig.~\ref{fig:masersPanels} show the distributions (as a function of the $X$ positions) of the $V_R$ and $V_\theta$ velocity components, respectively ($V_{R}=p_R$, see Section~\ref{sec:Hamiltonian}). The Galactic radii $R$ and azimuthal angles $\varphi$ were obtained directly from the positions and parallaxes of the sources, as well as their uncertainties.


\section{Galactic model}
\label{sec:model}

To study the corotation resonance, we make use of a two-dimensional model which represents the Galactic mid-plane. The model introduces the Galactic gravitational  potential which consists of an axisymmetric contribution due to the bulk distribution of matter in the Galactic main components, plus perturbations due to the spiral arms.
Despite being aware of the central Galactic bar, we do not consider a bar component in our calculations.
We performed preliminary tests adding a Dehnen-like bar \citep{dehnen2000AJ} in the center of the Galaxy, with the adopted spiral structure of this paper, and the result was that the effect of the bar was negligible at the solar radius, unless we use an unrealistic bar strength, which would distort the observed spiral arms. We therefore conclude that its effects on the stellar motion at the Sun's distance may be neglected in a first approximation.

\subsection{Rotation curve and axisymmetric potential}\label{sec:rotcur}
The axisymmetric potential $\Phi_0(R)$ is defined by the rotation curve $V_{\rm rot}(R)$ via the relation
\begin{equation}
\partial\Phi_0/\partial R = V^2_{\rm rot}(R)/R\,,
\label{eq:axisymmetric}
\end{equation}
where $R$ is the galactocentric radius.
We adopt a realistic rotation-curve model of the Milky Way  based on published observational data \citep{clemens1985ApJ,fichBlitzStark1989ApJ,reidEtal2014ApJ}. We assume the Galactocentric distance of the Sun \mbox{$R_0=8.0$\,kpc} and the velocity \mbox{$V_{0}=230$\,km\,s$^{-1}$} of the local standard of rest (LSR) (see \citealp{michtchenkoVieiraBarrosLepine2017AA} and references therein). We fit the observational data by the sum of two exponentials in the form (see \citealt{michtchenkoVieiraBarrosLepine2017AA} for details of the data used and the fitting procedure)
\begin{eqnarray}
 V_{\rm rot}(R) &=& 298.9\,\exp\left(-\frac{R}{4.55}-\frac{0.034}{R}\right)  \nonumber\\
 & & + 219.3\,\exp\left[-\frac{R}{1314.4}-\left(\frac{3.57}{R}\right)^2\right] ,
 \label{eq:Vrot}
\end{eqnarray}
\noindent
with the factors multiplying the exponentials given in units of kilometers per second and the factors in the arguments of the exponentials given in kiloparsecs.
We use  the trapezium rule with adaptive step to solve numerically Eq.~(\ref{eq:axisymmetric}) and obtain the axisymmetric gravitational potential $\Phi_0(R)$ adopting the ``numerical infinity'' condition $\Phi_{0}(1000\,{\rm kpc})=0$. It is worth noting that our approximation needs no assumptions on what components of the Galaxy (stellar or gaseous matter, dark matter, etc) are effectively contributing to the axisymmetric potential at each radius.

\subsection{Spiral arms potential}

Observations of spiral tracers favour four-arm models \citep[which conclude that these structures are most likely long lived]{houHan2009AA, houHan2014AA}, as corroborated by tangent-line data \citep{vallee2016AJ}. In this work, the adopted spiral  pattern consists of four logarithmic spirals which are described as valleys in the gravitational potential, with Gaussian cross-section (the half width, in the direction perpendicular to the spiral, is equal to 0.95\,kpc). This description of the arms was first introduced by \citet{junqueiraEtal2013AA} and has been used successfully in recent works \citep{barrosLepineJunqueira2013MNRAS, liEtal2016ApJ, michtchenkoVieiraBarrosLepine2017AA}. Concerning the spiral loci, observational and theoretical studies assume distinct values for the pitch angle, as summarized in \cite{vallee2015MNRAS}. Based on a comparison between different determinations of the pitch angle, that paper proposes $-13^{\circ}.0  \pm 0^{\circ}.6$ (with the negative sign, which we adopt hereafter as just a matter of convention). This value is in accordance with that derived by \citet{bobylevBajtkova2013AstL} ($-13^{\circ}.7$) via the analysis of maser data, but notice that slightly different values have been suggested based on different tracers \citep{houHan2014AA}. In this work, we adopt a pitch angle of $-14^{\circ}.0$; this value adjusts well the observed positions of the masers sources with respect to the spiral arms (Fig.~\ref{fig:bracosXY}). The spiral potential is given by

\begin{equation}
\label{eq:H1gaussian}
 \Phi_{\rm s}(R,\varphi) = -K(R)\times \zeta_0\,R\,e^{-\frac{R^2}{\sigma^2}[1-\cos(m\varphi-f_m(R))]-\varepsilon_s R},
\end{equation}
where $m=4$ is the number of arms and $\varphi$ is the azimuthal angle in the frame rotating with anglular velocity $\Omega_p$. In order to cut off arms effects inside the central region ($R < 3$\,kpc), we truncate the spirals by  multiplying the spiral amplitude by a function $K(R)$ of the form \citep{contopoulosGrosbol1988AA}
$$
K(R) = 0.5\,\big[ 1+\tanh\left[b_1\,(R-b_2)\right]\big].
$$
The values of the parameters for the spiral perturbation used in this paper are given in Table~\ref{tab:parameters}; 
in particular, the maximum value of the ratio 
$\big|(\partial\Phi_s/\partial R)/(\partial\Phi_0/\partial R)\big|$ is about 4\% (see also \citealp{michtchenkoVieiraBarrosLepine2017AA}).

The spiral structure rotates rigidly with pattern speed $\Omega_p$. The shape function $f_m(R)$ is given by
\begin{equation}\label{eq6}
f_m(R) = \frac{m}{\tan(i)}\ln{(R/R_i)}+ \gamma
\end{equation}
where $i$ is the spiral pitch angle; $R_i$ is a reference radius and $\gamma$ is a phase angle, whose values define the orientation of the spirals in the chosen reference frame; in this paper, $R_i=8$\,kpc and $\gamma=237^\circ.25$ (for $i=-14^\circ$; for the choice of the reference frame, see Section~\ref{energy_levels}). The black spiral curves in Fig.~\ref{fig:bracosXY} correspond to the loci of the main spiral arms in our model; they are the azimuthal minima of the spiral potential $\Phi_{\rm s}(R,\varphi)$  (\ref{eq:H1gaussian}). The gray shading represents the intensity of the Gaussian-shaped spiral potential.

The nominal (approximate) value of the corotation radius $R_{CR}$  is obtained from the relation
\begin{equation}
V_{\rm rot}(R_{CR})=\Omega_p\,R_{CR}.
\end{equation}
With the adopted value of $\Omega_p=28.5$\,km\,s$^{-1}$\,kpc$^{-1}$ (see Section~5), we obtain $R_{CR}=8.06$ kpc.



\begin{deluxetable}{lclc}
\tabletypesize{\footnotesize}
\tablecaption{Adopted spiral arms parameters.\label{tab:parameters}}
\tablewidth{0pt}
\tablenum{2}
\tablecolumns{4}
\tablehead{
\colhead{Parameter} & \colhead{Symbol} & \colhead{Value} & \colhead{Unit}
}
\startdata
Solar radius          & $R_\odot$           & 8.0         & kpc  \\
LSR velocity          & $V_0$               & 230       & km s$^{-1}$  \\
Number of arms        & m                   & 4         & -   \\
Pitch angle           & i                   & -14$^\circ$& -   \\
Arm width             & $\sigma \sin {\rm i}$ & 1.94       & kpc \\
Scale length          & $\varepsilon_s^{-1}$& 4.0       & kpc  \\
Spiral pattern speed  & $\Omega_p$          & 28.5        & km\,s$^{-1}$\,kpc$^{-1}$\\
Spiral amplitude& $\zeta_0$           & 200.0      & km$^2$\,s$^{-2}$\,kpc$^{-1}$\\
Reference radius      & $R_i$               & 8.0        & kpc\\
Cutoff coefficient 1  & $b_1$               & 2.5      & kpc$^{-1}$ \\
Cutoff coefficient 2  & $b_2$               & 2.0       & kpc \\
\enddata
\end{deluxetable}


\section{Hamiltonian topology: energy levels, equilibria and spiral arms}
\label{sec:Hamiltonian}

The Hamiltonian which describes the stellar dynamics in the galactic mid-plane is given by the sum of the axisymmetric $\Phi_0(R)$ and the spiral $\Phi_{s}(R,\varphi)$ potentials as
\begin{equation}\label{eq:Hamil}
{\mathcal H}(R,\varphi,p_R,L_z)= {\mathcal H_{0}}(R,p_R,L_z) + \Phi_{s}(R,\varphi)\,
\end{equation}
where ${\mathcal H_{0}}$ is the unperturbed component of the Hamiltonian in the rotating reference frame, given by Jacobi's integral
\begin{equation}\label{eq:H0}
{\mathcal H_{0}}(R,p_R,L_z)= \frac{1}{2}\left[ p_R^2 + \frac{L_z^2}{R^2}\right] + \Phi_0(R) - \Omega_p L_z \,.
\end{equation}
Here, $p_R$ and $L_z$ are the canonical momenta conjugated to $R$ and $\varphi$, respectively.

\subsection{Stationary solutions}

The stationary solutions of the Hamiltonian flow (\ref{eq:Hamil}) are given by the set of equations \citep{michtchenkoVieiraBarrosLepine2017AA}
\begin{eqnarray}
\frac{\partial \Phi_0(R)}{\partial R} +\frac{\partial \Phi_s(R,\varphi)}{\partial R} &=&\frac{L_z^2}{R^3}, \label{eq:H00-1}\\
m\varphi&=&\varphi_0+f_m(R) \label{eq:H00-2}\\
p_R &=& 0, \label{eq:H00-3}\\
L_z &=& \Omega_pR^2, \label{eq:H00-4}
\end{eqnarray}
where \mbox{$\varphi_0=\pm n\,\pi$} and \mbox{$n=0, 1, ...$}. The symmetry of this  problem is $2\,\pi / m$ (which in our case is a four-fold symmetry).

\subsection{Energy levels on the $X$--$Y$ plane}
\label{energy_levels}

The topology of the Hamiltonian (\ref{eq:Hamil}) is visualized by plotting the energy levels on a representative plane. The representative plane is defined as a plane of initial conditions chosen in such a way that all possible configurations of the system are included, and thus all possible regimes of motion of the system under study can be represented on it.

The plane ($X=R\cos\varphi$,\,$Y=R\sin\varphi$) is widely used in the literature to present the modeled structure of the Galaxy and we start our study with this conventional choice (see Fig.~\ref{fig:bracosXY}). First, we fix the reference axis $X$ ($\varphi=0$) in such a way that the Sun's azimuthal coordinate is $\varphi = 90^\circ$, placing the Sun on the $Y$-axis at $R=8.0$\,kpc. The orientation of the spiral arms given by Eq.\,(\ref{eq:H00-2}) on the $X$--$Y$ plane is defined by the value of the free parameter $\gamma$ in the expression (\ref{eq6}). We choose $\gamma$ such that the Sun (located at $R=8.0$\,kpc and $\varphi=90^\circ$) is 1\,kpc distant from  the Sagittarius arm locus defined by $\varphi_0=0$ in Eq.\,(\ref{eq:H00-2}); thus, we obtain $\gamma=237^\circ.25$. 

For the obtained value of $\gamma$ (since we use the positive value for the pitch angle in Fig.~\ref{fig:bracosXY}, we must reflect these obtained $\gamma$--values with respect to the \mbox{$Y$-axis}), the spiral arms loci of the system are calculated from Eq.\,(\ref{eq:H00-2}) with even values of $n$ and plotted on the $X$--$Y$ plane (Fig.~\ref{fig:bracosXY}).
The spiral potential $\Phi_{\rm s}(R,\varphi)$ in Eq.\,(\ref{eq:H1gaussian}) is plotted using a gray-level scale: the lowest black level matches the spiral arms associated to the minima of the potential, as the color bar shows.

The effective potential of the system is given by
\begin{equation}\label{eq:effectivepotential}
\Phi_{\rm eff}(R,\varphi) = \Phi_0(R) + \Phi_{\rm s}(R,\varphi) - \frac{1}{2}\Omega_p^2R^2
\end{equation}
and is equal to the total energy of the system given by Eq.\,(\ref{eq:Hamil}), subject to the conditions (\ref{eq:H00-3}) and (\ref{eq:H00-4}). It is plotted by the level curves in Fig.~\ref{fig:bracosXY}. The corotation domains appear as banana-like regions (red curves) located between the spiral arms: the maximal energy  stationary solutions (libration centers) are deeply inside the corotation islands, while the minimal energy  stationary solutions lie on the spiral arms. There are four libration centers whose positions on the $X$--$Y$ plane are given by the corotation radius $R_{CR}$ and the corotation angles $\varphi_{CR}+k\pi/2$. The corotation coordinates are solutions of the set of the equilibrium conditions given by Eqs.\,(\ref{eq:H00-1})--(\ref{eq:H00-4}) with $n$ odd and are strongly dependent on the assumed value of the pattern speed $\Omega_p$.
In our model, for the parameters from Table~\ref{tab:parameters}, the Sun evolves inside the corotation island around the libration center with coordinates $R_{CR}=8.06$\,kpc and $\varphi_{CR}=76^\circ$, defined as the local corotation zone in this paper.

We must remark that the $X$--$Y$ plane is not a representative plane for the system, as mentioned in \citet{michtchenkoVieiraBarrosLepine2017AA}. The conditions (\ref{eq:H00-3}) and (\ref{eq:H00-4}), imposed to the Hamiltonian (\ref{eq:Hamil}) in order to obtain the effective potential, are too restrictive. We must therefore obtain another plane which would represent the full dynamics inside the corotation zone. This subject is postponed to Section~\ref{sec:map}.

\begin{figure}
\begin{center}
\epsfig{figure=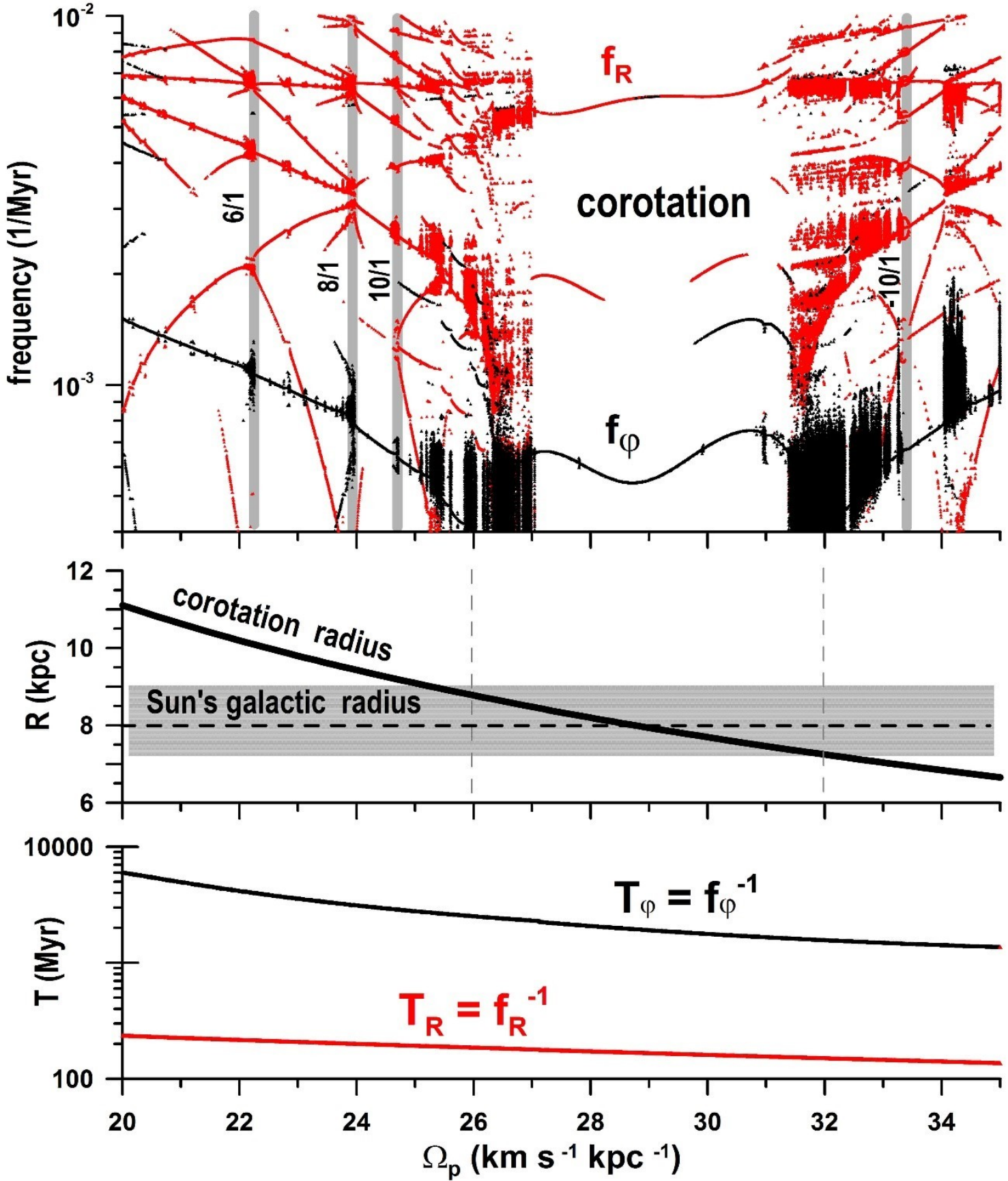,width=0.99\columnwidth ,angle=0}
\caption{\emph{Top}: Proper frequencies $f_R$ (red) and $f_\varphi$ (black) (in logarithmic scale), their harmonics and linear combinations, as functions of the spiral pattern speed $\Omega_p$, for the maser \# 14 from Table~\ref{tab:masers} in Appendix.
The local corotation zone lies in the $\Omega_p$--range between 27 and 31\,km\,s$^{-1}$\,kpc$^{-1}$.  Some Lindblad resonances are indicated by vertical lines and the corresponding ratio.
\emph{Middle}: Evolution of the corotation radius as a function of $\Omega_p$; the shaded horizontal strip covers the range of the current radii of the Local arm's maser objects from Table~\ref{tab:masers} in Appendix. The vertical dashed lines delimit the estimated $\Omega_p$--range of the corotation resonance.
\emph{Bottom}: Characteristic radial ($T_R$) and azimuthal ($T_\varphi$) periods of orbits in the vicinity of the stable corotation center, as functions of $\Omega_p$.
}
\label{fig:powerSpec}
\end{center}		
\end{figure}




\section{Estimate of the spiral pattern speed}
\label{sec:patternspeed}

The pattern speed $\Omega_p$ and the galactic rotation curve determine the corotation radius. Assuming that the motions of the Sun and masers are close to or inside the corotation resonance, we constrain the possible values of $\Omega_p$. In order to estimate the range of the $\Omega_p$-values, we analyse the dynamics of the Local arm tracers presented in Fig.~\ref{fig:bracosXY} (masers as red crosses; see Table~\ref{tab:masers} in Appendix for their identifications, positions and velocities). First, we construct the dynamical power spectrum (see the Appendix and \citealp{michtchenkoVieiraBarrosLepine2017AA}) parameterized by the values of $\Omega_p$, for maser \#14 from Table~\ref{tab:masers} in Appendix; this object was chosen from the middle of the distribution of the maser sample in phase space (see Fig.~\ref{fig:masersPanels}). The dynamical power spectrum in Fig.~\ref{fig:powerSpec}\,top shows the evolution of the proper frequencies $f_R$ (red) and $f_\varphi$ (black), their harmonics, and linear combinations with the spiral pattern speed $\Omega_p$. The frequencies $f_R$ and $f_\varphi$ were calculated by analysing the time evolution of $R(t)$ and $L_z(t)$ of the maser's orbit, respectively.  The smooth evolution of the frequencies with $\Omega_p$ is characteristic of regular motion, while the erratic scattering of the points is characteristic of chaotic motion. The corotation zone extends in the range of the $\Omega_p$--values from 27 to 31\,km\,s$^{-1}$\,kpc$^{-1}$, in this particular case, for which the object \#14 evolves inside the stable region of the corotation resonance. The corotation zone is delimited by the thick layers of chaotic motion.

We could extend this analysis to other objects from our sample; however, we prefer to proceed in a different way. Owing to the fact that, inside the corotation resonance, the radial coordinate of the masers must oscillate around the value of the corotation radius \citep{michtchenkoVieiraBarrosLepine2017AA},  we calculate $R_{CR}$ using our full Hamiltonian model (see Eqs.\,(\ref{eq:H00-1})--(\ref{eq:H00-4})), for various values of the pattern speed, and  compare its values with the current radii of the Local arm objects. The family of corotation radii as a function of $\Omega_p$ is shown in Fig.~\ref{fig:powerSpec}\,middle, together with the radial distances of the Local arm objects  distributed in a horizontal gray strip. The current radii of the orbits of these objects vary between 7.1\,kpc and 9.2\,kpc (see Table~\ref{tab:masers} in Appendix).  We then obtain approximately  the limits to the possible $\Omega_p$--values between 26 and 32\,km\,s$^{-1}$\,kpc$^{-1}$, which are in good agreement with observational determinations \citep{diasLepine2005ApJ,gerhard2011MSAIS}.
Note that these limits correspond to the set of the spiral arms parameters from Table~\ref{tab:parameters}.

In this work, we adopt $\Omega_p=28.5$\,km\,s$^{-1}$\,kpc$^{-1}$, a value which is situated in the middle of the interval defined by maser \#14. Together with the other adopted galactic parameters (see Table~\ref{tab:parameters}), this value represents well our basic hypothesis, which presumes that the Local arm lies inside the local corotation zone. In other words, with $\Omega_p=28.5$\,km\,s$^{-1}$\,kpc$^{-1}$, the majority of maser sources of the Local arm, as well as the Sun, lie inside the local corotation zone (see Sections~\ref{dep_on_Omegap} and~\ref{sec:map}). It is worth emphasizing that  small changes in the adopted value of $\Omega_p$ will alter quantitative features of the system but will preserve the qualitative dynamics \citep{michtchenkoVieiraBarrosLepine2017AA}, keeping unaltered our final results.

\subsection{Dependence of the solar orbit on $\Omega_p$}
\label{dep_on_Omegap}

\begin{figure}
\begin{center}
\epsfig{figure=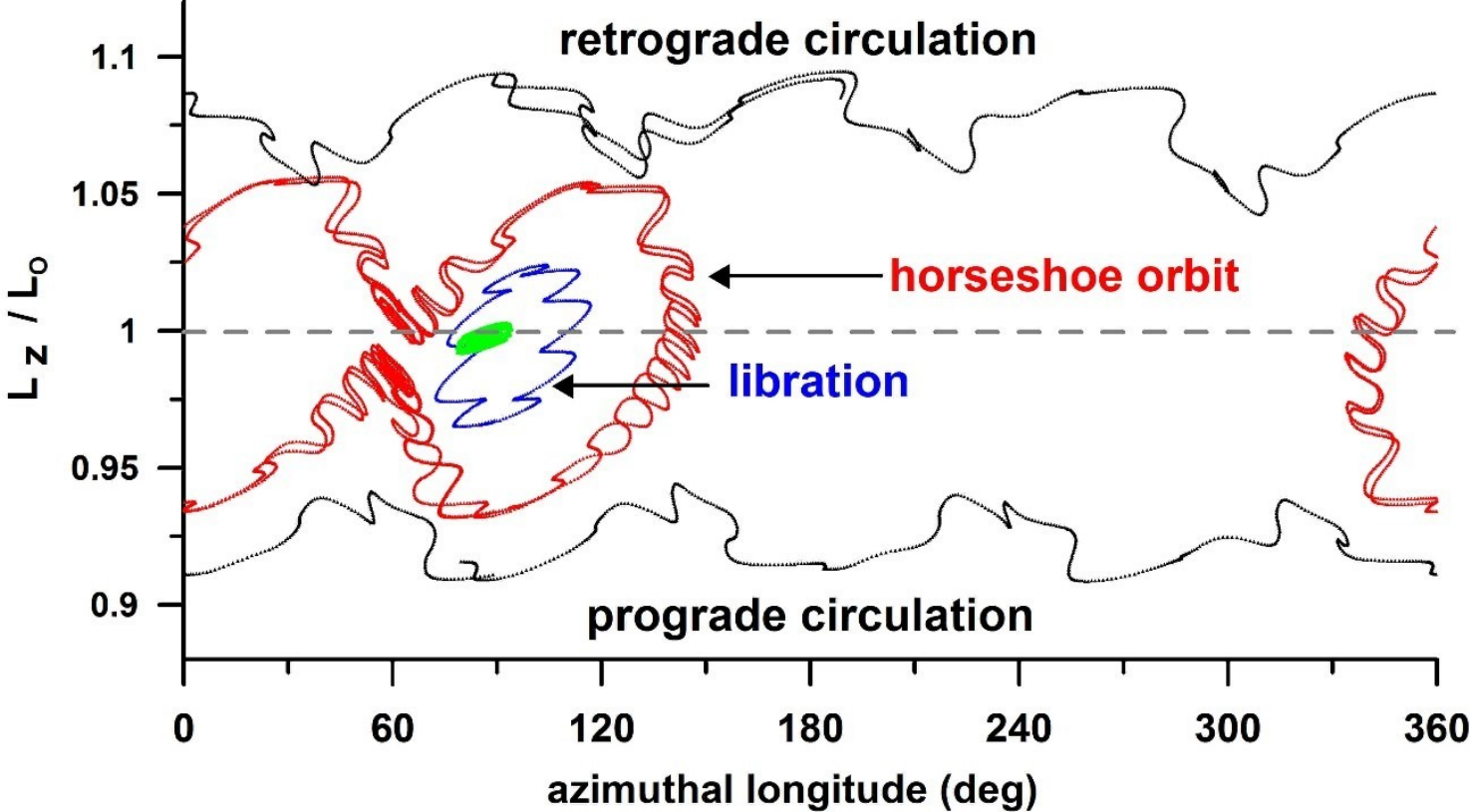,width=0.99\columnwidth ,angle=0}
\caption{Family of the Sun's orbits parameterized by different values of the pattern speed $\Omega_p$, on the plane $\varphi$--$L_z/L_0$, where $L_0$ is the angular momentum of the corresponding corotation center (the solar coordinates are given in Table~\ref{tab:masers} in Appendix). Two orbits (black) are circulating: prograde, for $\Omega_p=24.0$ (in units of km\,s$^{-1}$\,kpc$^{-1}$), and retrograte, for $\Omega_p=30.0$. For two $\Omega_p$--values, $25.0$ and $29.0$, the orbits are horseshoe-like orbits (red), enclosing two resonant islands and oscillating around  $L_z/L_0=1$. For two $\Omega_p$--values, $26.0$ and $28.0$, the orbits (blue) are librating inside the local corotation zone. Finally, for $\Omega_p=27.0$, the orbit (green) librates around  $L_z/L_0=1$ with a very small amplitude of oscillation, indicating that the Sun is located very close to the corotation center.
}
\label{fig:orbitSun}
\end{center}
\end{figure}

The qualitative aspects of the solar orbit are strongly dependent on the value of $\Omega_p$; even for reasonable values, the Sun may be either inside or outside the corotation resonance. Fig.~\ref{fig:orbitSun} shows the projection of the Sun's orbit in the $\varphi-L_z/L_0$ plane, for different values of $\Omega_p$, in the interval 24--30 \,km\,s$^{-1}$\,kpc$^{-1}$. Here, $L_0$ is the equilibrium value of $L_z$, given by $\Omega_p R_{CR}^2$ (that is, evaluated for the stable fixed point of the Hamiltonian which is next to the Sun). For $\Omega_p$--values 24 and 30 \,km\,s$^{-1}$\,kpc$^{-1}$, the solar orbit is circulating (prograde and retrograde, respectively, with respect to the rotating frame); for 25 and 29 \,km\,s$^{-1}$\,kpc$^{-1}$ the Sun is on a horseshoe orbit\footnote{We use the term \emph{horseshoe orbit} in the context of Celestial Mechanics \citep{murray1999solar}; note that some authors use this term to refer to orbits encompassing only one stable fixed point of the Hamiltonian, e.g. \cite{sellwoodBinney2002MNRAS.336..785S}.} (encompassing two stable fixed points of the Hamiltonian), while for 26, 27, and 28 \,km\,s$^{-1}$\,kpc$^{-1}$ the Sun librates around a single stable fixed point, which defines the local corotation zone. We therefore estimate that, in order to have the Sun in the corotation resonance, the value of $\Omega_p$ must be in the region 25--29 \,km\,s$^{-1}$\,kpc$^{-1}$, what is in accordance with our choice $\Omega_p=28.5$\,km\,s$^{-1}$\,kpc$^{-1}$.
With this adopted value for $\Omega_p$ and the rotation curve used in this work (see Section 2.1), the nominal value of the corotation radius is obtained as $R_{CR}=8.06$ kpc.

\newpage

\section{Integration of masers orbits and the dynamical map}
\label{sec:map}

\begin{figure}
\begin{center}
\epsfig{figure=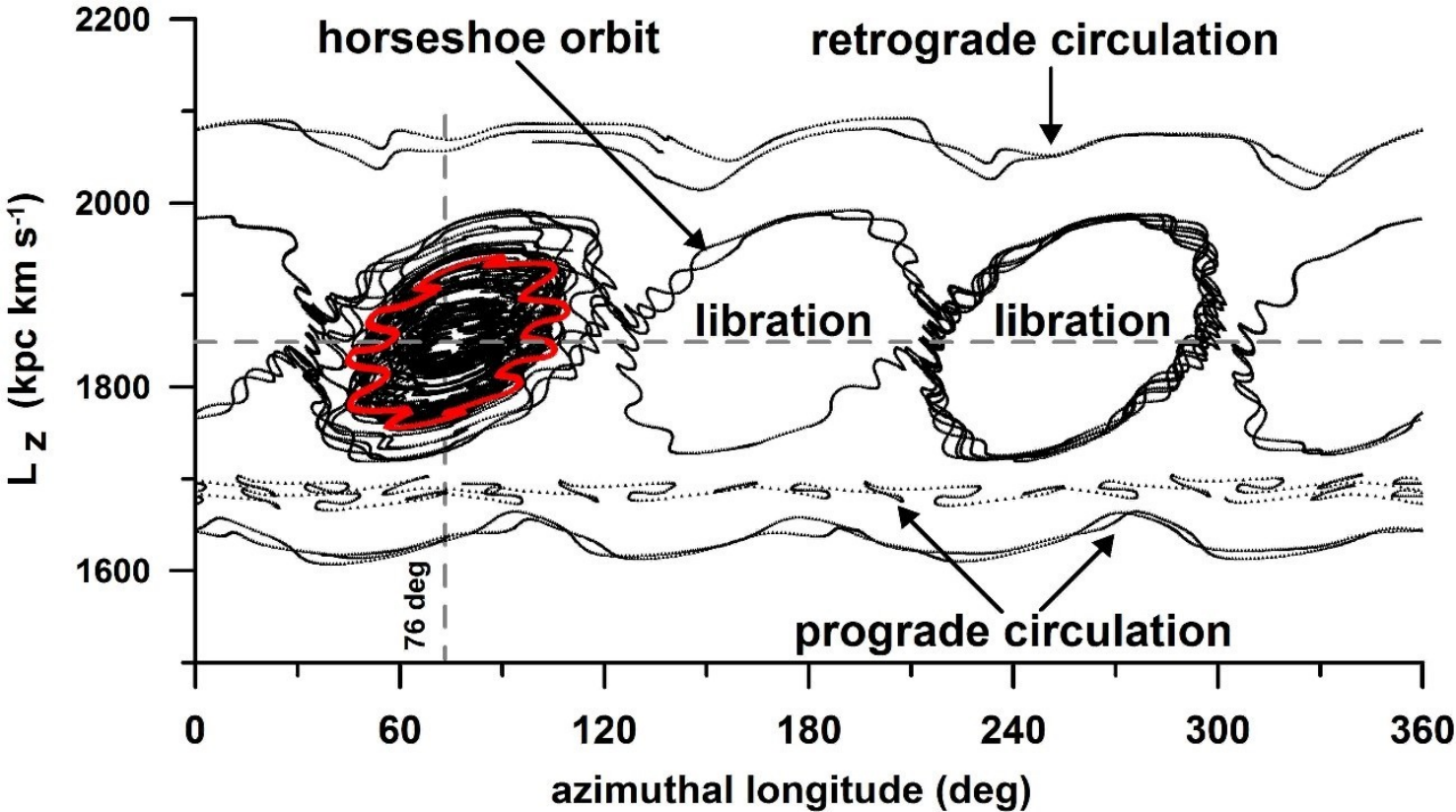,width=0.99\columnwidth ,angle=0}
\caption{Examples of the orbits of the maser sources from the Local arm (black) and the orbit of the Sun (red), projected on the plane $\varphi$--$L_z$. The initial conditions of the Sun and maser objects are given in Table \ref{tab:masers}, while the parameters adopted in the galactic model are given in Table \ref{tab:parameters}. The equilibrium $L_z$--value is shown by the horizontal dashed line, at $1851.5$\,kpc\,km\,s$^{-1}$.
}
\label{fig:phiLzOrbits}
\end{center}
\end{figure}

We integrated numerically the orbits of the 47 maser sources which trace the Local arm. The values of the parameters used in our Hamiltonian model are given in
Table~\ref{tab:parameters}, while the initial conditions of the masers, i.e. their current positions and velocities, are given in Table~\ref{tab:masers} in Appendix. An initial analysis was performed by inspecting the projection of the orbits on the $\varphi-L_z$ plane (Fig.~\ref{fig:phiLzOrbits}). We find that 40 objects from our masers sample, as well as the Sun, oscillate around the equilibrium value of $L_z$ (dashed line in Fig.~\ref{fig:phiLzOrbits}); these masers are shown as filled red circles in the panels of Fig.~\ref{fig:masersPanels}. This librating behaviour indicates that these objects are trapped inside the corotation resonance. Among the librating maser sources, the motion of 37 of them (and of the Sun) is confined to the one resonant island of stability centered at $\varphi_{CR}=76^{\circ}$. The trajectories of the three other maser sources (\#5, \#7 and \#16 in Table~\ref{tab:masers} in Appendix) show a complicated structure known as ``horseshoe orbit'' (see footnote on Section~\ref{dep_on_Omegap}), characteristic of the objects which leave the libration island of origin  and start to encompass the other (four in this case) resonant islands, but  never perform
cycles of $360^\circ$. The other 7 maser sources do not belong to the local corotation zone in our model; nevertheless, they are very close to this resonance. Their orbit's projections do not cross the equilibrium value of $L_z$ and therefore are necessarily circulating: the circulation is prograde (in the rotating frame) for lower velocities (5 objects) and retrograde for higher velocities (2 objects).

\begin{figure}
\begin{center}
\epsfig{figure=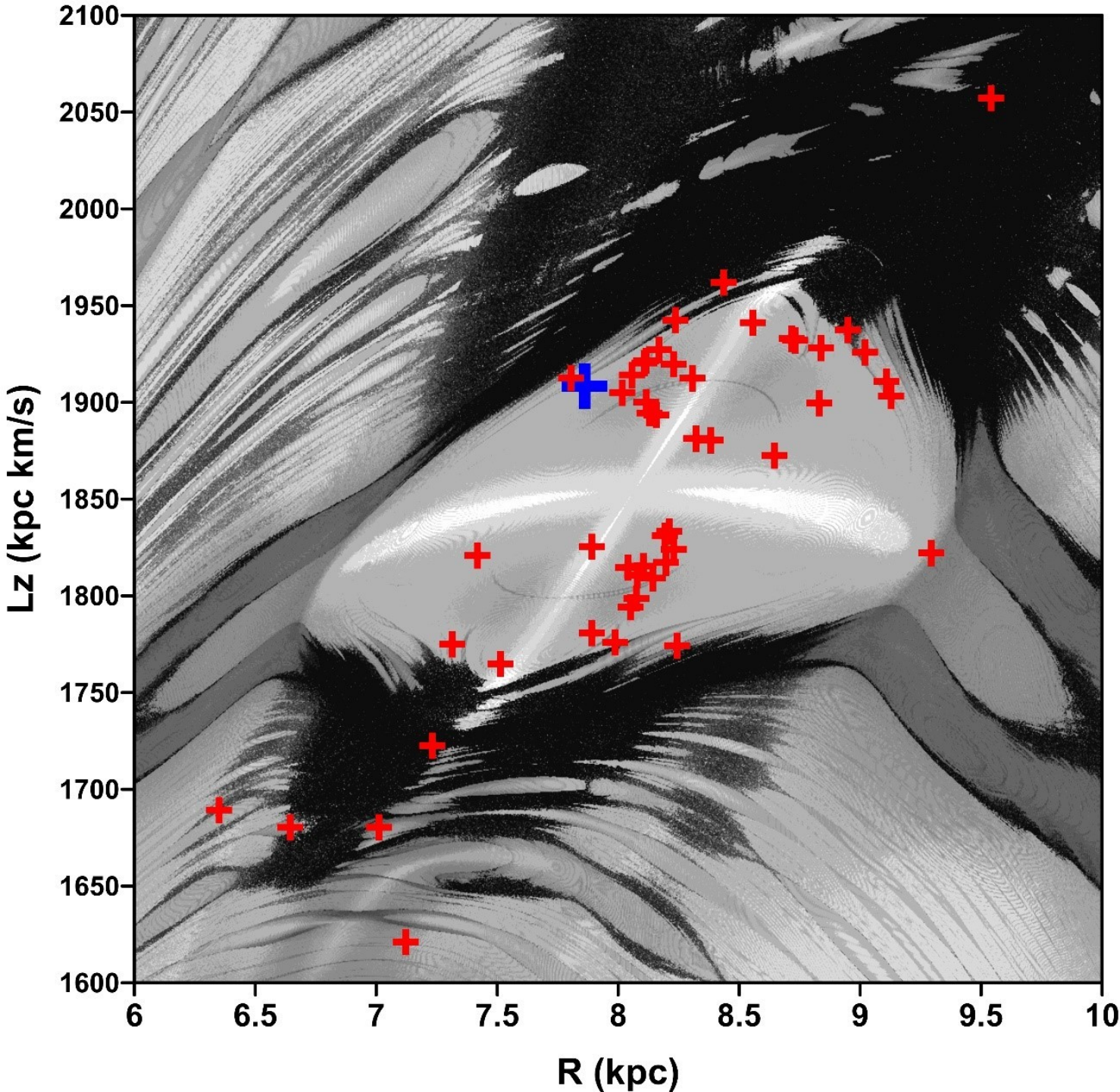,width=0.99\columnwidth ,angle=0}
\caption{Dynamical map on the representative $R$--$L_z$ plane, calculated for the equilibrium conditions $p_R=0$ and $\varphi_{CR}=76^\circ$. The central gray-white region represents the local corotation zone. The Sun is depicted by a blue cross. The locations of the 47 maser sources are shown by red crosses. The positions of the Sun and masers on the map are obtained by propagating their current positions and velocities, until each object
crosses the representative plane.
}
\label{fig:dynamicalMap}
\end{center}
\end{figure}

Dynamical maps (see \citealp{michtchenkoVieiraBarrosLepine2017AA} and the Appendix) are constructed numerically using the information provided by the spectral number $N$, to characterize the phase-space structure and identify the precise location of the corotation and other resonances. The plane ($X=R\cos\varphi$;\,$Y=R\sin\varphi$) described above does not represent the whole set of initial conditions since it is limited to the $L_z$--values given by Eq.\,(\ref{eq:H00-4}). \cite{michtchenkoVieiraBarrosLepine2017AA} suggested the plane $R$--$V_\theta$, with $p_R=0$ and $\varphi=\varphi_{CR}$, as a good representative plane. In this work, we choose the representative plane $R$--$L_z$ of initial conditions (spanning the whole corotation region and encompassing the orbits of all Local-arm maser sources; for details, see \citealp{michtchenkoVieiraBarrosLepine2017AA} and the Appendix), with $p_R=0$ and $\varphi=\varphi_{CR}$, which is covered by a fine grid. We construct a dynamical map over this representative plane of initial conditions (Fig.~\ref{fig:dynamicalMap}) by labeling each initial condition on this grid by its corresponding spectral number $N$, calculated from the time series of the radial coordinate $R(t)$ for each orbit. We associate with this spectral number a logarithmic gray-scale varying from white (\mbox{$N=1$}) to black ($N$ maximum). Here we choose $N=100$ for the maximum value of $N$. All orbits with $N>100$ are also labelled black. Resonances appear as chains of regular islands in this map, surrounded by regions of chaos.

A thorough analysis of the phase-space structure around corotation is presented in the dynamical map of Fig.~\ref{fig:dynamicalMap}. It shows the stable domain of the corotation resonance (the central light-colored region in Fig.~\ref{fig:dynamicalMap}) delimited by the layers of chaotic motion (dark-colored regions), which are associated with the separatrix of the resonance. The solar orbit, as well as all orbits of the maser sources belonging to the Local arm, were time propagated numerically, until crossing the representative plane and the corresponding $R$ and $L_z$ coordinates of each object were then plotted on the dynamical map (red symbols in Fig.~\ref{fig:dynamicalMap} for maser sources, blue cross for the Sun). We confirm that the majority of the maser orbits lie inside the corotation island of stability and this is also valid for the solar orbit.

\section{Open clusters}
\label{clusters}

In the present work, we also study the effects of the local corotation zone, which generates the Local arm, on the orbital properties of a sample of Open Clusters (OCs) younger than 100 Myr. The advantage of using a sample of young objects is that they could have moved to only little distances from their birthplaces, so that their current positions in the Galactic plane must resemble approximately the overall picture of the spiral arms in regions not too far from the Sun. Unlike the masers, we do not have a pre-established list of OCs associated with the Local arm, but we can characterize the objects that are most likely under the influence of the local corotation zone based on their locations in the phase space of the galactic system.

The data for OCs were retrieved from the New Catalogue of Optically Visible Open Clusters and Candidates\footnote{http://www.wilton.unifei.edu.br/ocdb/}, published by \cite{diasEtal2002AA}. We selected a sample of objects with determined distances and measured heliocentric radial velocities and proper motions. The Galactic radii $R$, azimuthal angles $\varphi$, and Galactocentric velocities $V_R$ and $V_\theta$ of the OCs were calculated in the same manner as for the maser sources, as described in Section~\ref{sec:masers}.

\begin{figure}
\begin{center}
\epsfig{figure=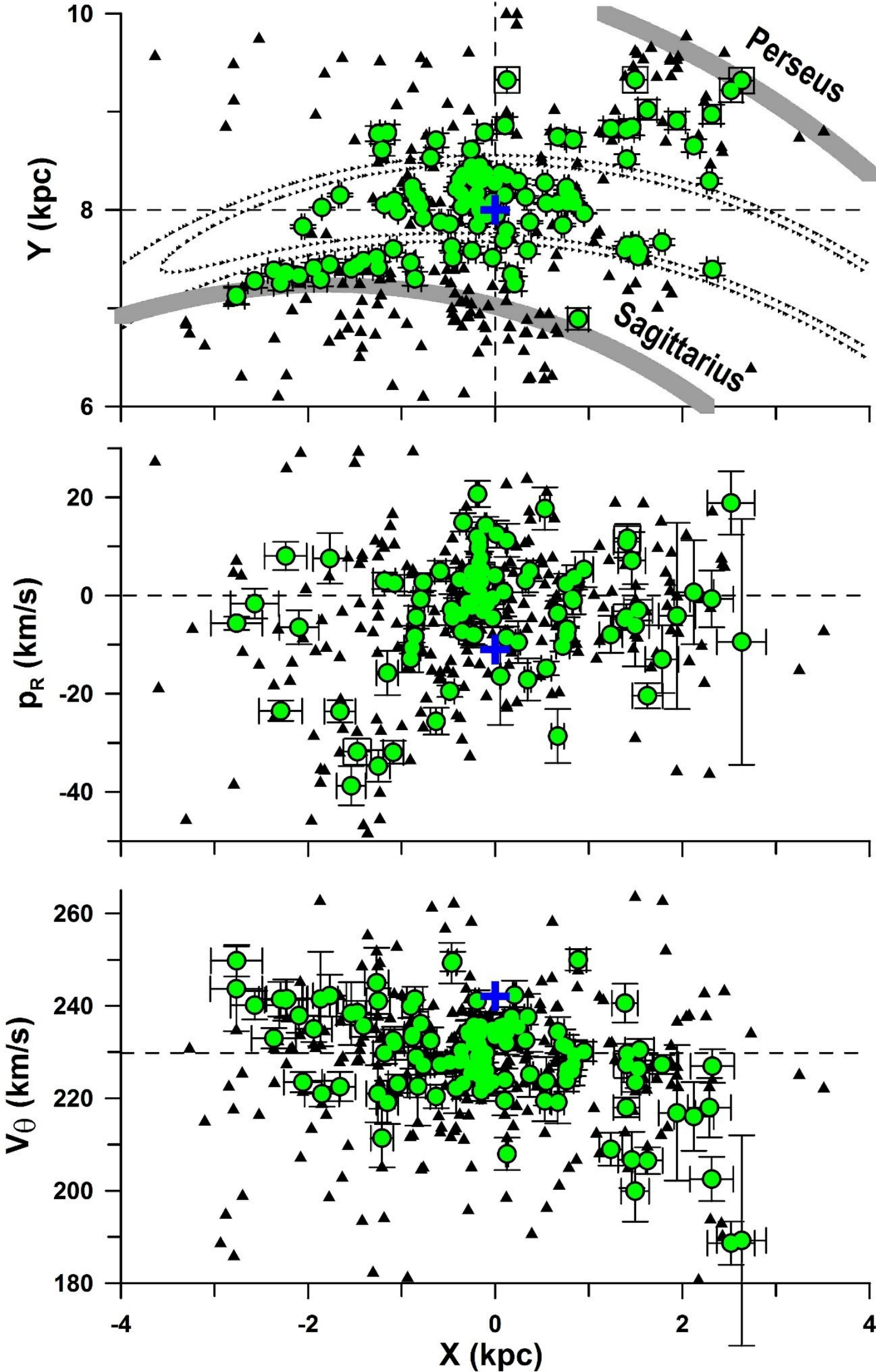,width=0.99\columnwidth ,angle=0}
\caption{
 Same as Fig.\ref{fig:masersPanels}, except for OCs (triangles and green circles).
}
\label{fig:openclusters}
\end{center}
\end{figure}

We now discuss the main results obtained from the sample of OCs. The positions of the young OCs (ages $<100$ Myr), selected as described above, are presented in Fig.~\ref{fig:openclusters}. The OCs marked as green circles (with error bars) are those trapped in the local corotation zone, as revealed by the numerical integration of their orbits for a time interval of 5 Gyr. The figure shows that a large portion of the young OCs contained in the banana-like stability zone are trapped in it.
The fact that the banana-like region is not uniformly filled with OCs is probably an observational selection effect, as the OCs have usually been discovered by visual inspection of photographic plates. The distribution of OCs's distances from the Sun presents a strong decrease in the number of objects beyond 2 kpc (see Fig. 1 of \citealp{wuEtal2009MNRAS}).

Curiously, Fig.~\ref{fig:openclusters}\,top shows a number of OCs which are trapped in the local corotation zone, but situated very close to the Sagittarius-Carina arm.
The apparent scattering in the distribution of trapped OCs around the banana-like region in Fig.~\ref{fig:openclusters}\,top is due to the spread in velocities of the sample: The corotation zone is a domain in the four-dimensional phase space, whereas the banana-like region represents its projections on the two-dimensional $X$--$Y$ plane subject to conditions (\ref{eq:H00-3}) and (\ref{eq:H00-4}). We show in the middle and bottom panels in Fig.~\ref{fig:openclusters} the radial momenta $p_R$ and tangent velocities $V_\theta$ of the OCs presented in Fig.~\ref{fig:openclusters}\,top, together with the corresponding error bars. Comparing the three panels, we see that the objects whose projection on the $X$--$Y$ plane deviate from the banana-like region have, in general, velocities which deviate significantly from their equilibrium values (dashed lines). If the current positions and velocities were propagated until reaching the conditions (\ref{eq:H00-3})--(\ref{eq:H00-4}), which define the effective potential, they would lie inside the banana-like region of Fig.~\ref{fig:bracosXY}. The difference between the equilibrium values of $p_R$ and $V_\theta$ and the corresponding OC values is therefore the cause of the observed scattering in positions (some of them outside the banana-like region).
It is expected, therefore, that there are more objects trapped in the local corotation zone than those of the Local arm; the OCs are an example of this phenomenon.

\section{Discussion}
\label{sec:discussion}


We have evidence that the Sun is located near the spiral corotation circle \citep{mishurovZenina1999AA,diasLepine2005ApJ,gerhard2011MSAIS, lepineEtal2011MNRAS}. This result imposes constraints on the spiral pattern speed, assuming a long-lived spiral structure.
The present study confirms that the above evidence is consistent with a four-arm model for the Galaxy, constrained by spiral-tracer observations.
For the adopted spiral pattern speed, the Sun is found to be trapped in one of the four corotation islands of stability,  evolving with a characteristic radial short period of $\sim$168\,Myr and an azimuthal long period of $\sim$1.7\,Gyr with respect to the main spiral arms (see Fig.~\ref{fig:powerSpec}\,bottom). The range of galactocentric distances swept by the Sun is about 7.5--9\,kpc. 

Many efforts dedicated to the observational study of the Local arm tracers \citep{houHan2009AA, xuEtal2013ApJ, bobylevBajtkova2014MNRAS, houHan2014AA, reidEtal2014ApJ, xuEtal2016Science} have shown that the Sun also lies close to the Local arm, a bright arm segment lying between the Sagittarius-Carina and Perseus arms \citep{houHan2014AA}. Therefore, it is reasonable to assume that the Local arm is linked someway to the corotation resonance.
Analysing the dynamics of the masers sources associated with HMSFRs of the Local arm, whose positions and velocities were measured only recently, we found that the majority of these objects evolves within the limits of the corotation stability zone.
Although the maser sources are associated with massive stars, less massive stars are also being formed, according to the Initial Mass Function.

We now know, based on the numerical integration of their orbits, that these stars will not escape from the Local arm for long times. In some way, we are witnessing the building up of the Local arm. This process naturally produces a high density of stars along the stability zone, an effect which was already predicted for the corotation resonance \citep{contopoulos1973ApJ, barbanis1976AA}. Moreover, recent simulations of gas subject to a 2-arm spiral potential and a central bar generate 4-arm gas density responses with additional local arms which evolve self-consistently \citep{liEtal2016ApJ}.
The parameters corresponding to their best-fit model provide the spiral corotation circle located relatively close to the solar radius (they adopt $R_0=8.3$\,kpc and $R_{CR}=9.1$\,kpc). It corresponds to $R_{CR}/R_0=1.096$, which is in agreement with the range determined by \citet{diasLepine2005ApJ}, $R_{CR}=(1.06\pm 0.08)R_0$. This indicates that the spiral potential used in those simulations generates a gaseous local arm close to the Sun and to the spiral corotation radius.
Also, recent magnetohydrodynamics gas simulations of a galactic-like disk under the influence of a spiral potential have shown that banana-shaped orbits appear near the corotation circle \citep{gomezEtal2013MNRAS}, indicating that the dynamics of the gas is also subject to the corotation resonance.
Thus, we can conclude that the Local arm is a direct outcome of the action of the spiral local corotation zone.
Based on this result, we propose a scenario for the formation and evolution of the Local arm, which is presented in detail in Section~\ref{sec:evolution}.

Our results come from a  simple model for the disk of the Galaxy, based on the best available data on the local structure. The conclusions are robust against details of the spiral structure, including small variations of the pattern speed and pitch angle (see \citealp{michtchenkoVieiraBarrosLepine2017AA}). They also suffer small influence from details of the rotation curve adopted, such as deviations from the adopted values of the Galactocentric solar radius $R_0$ and the circular velocity $V_0$ at $R_0$, or a tenuous velocity dip just outside the Solar radius, a feature proposed by various authors \citep[e.g.][]{sofueHonmaOmodaka2009PASJ, barrosLepineJunqueira2013MNRAS}. The most important parameter which defines the qualitative features obtained in our work is the spiral pattern speed $\Omega_p$, which defines the corotation radius.

Our model does not take into account mutual interactions between stars, as it is usually assumed in orbital galactic dynamics. However, trapped stars inside the corotation zone are close to each other most of the time, and in this case we may question whether mutual interactions are relevant as a second order effect. It is expected that this effect will perturb the motion of stars inside the corotation zone. A modelling of this additional potential is necessary, as first remarked in \citet{contopoulos1973ApJ}. Moreover, a full description of the system must consider the gas dynamics at the corotation resonance; these issues are outside the scope of the present work and deserve a deeper investigation.

Regarding the relation between the Local arm and the local corotation zone, we have now established the link between the former (an observational fact) and the latter (a dynamical property of the system). The Local arm lies inside the local corotation zone, and most probably was generated by this dynamical mechanism. On the other hand, there are also resonant objects which do not belong to the Local arm that we observe nowadays.
We show in Fig.~\ref{fig:openclusters} an example of this connection, namely the sample of young open clusters with their $X$--$Y$ positions in the Galactic plane. 
We find that, according to our model, the objects in green belong to the corotation resonance. Clearly, even being inside the corotation zone, some of these objects are not part of the Local arm as defined by any observational method based on its tracers \citep[e.g][]{xuEtal2013ApJ, bobylevBajtkova2014MNRAS, houHan2014AA,reidEtal2014ApJ, xuEtal2016Science}. Thus, we conclude that the Local arm is an apparent location of young resonant objects aligned with the corotation circle on the Galactic plane, whereas the extension of the corotation domain can be assessed solely in the whole four-dimensional phase space. Thus, the locations of the observed resonant objects are just the projections of the phase space positions on the $X$--$Y$ plane.

The consequences of our findings are manifold. Let us first consider the kinematics in the close neighborhood of the Sun. The explanation for the moving groups, or ``streaming motions'' observed in the velocity space of stars situated in the solar neighborhood (at distances smaller than about 100\,pc), has been uncertain for decades, and most often considered as disrupted open clusters. The Sirius, Coma Berenices, Hyades-Pleiades and Hercules ``branches'' have been well mapped in the UV plane by \cite{antojaEtal2008AA}. Only recently a consensus seems to have been reached, that the moving groups are due to some kind of Galactic resonance, involving the spiral arms and/or the central bar   \citep{skuljanEtal1999MNRAS, dehnen2000AJ, quillenMinchev2005AJ, antojaEtal2009ApJL, antojaEtal2010LNEA, antojaEtal2011MNRAS, famaeyEtal2012EPJWC}. It is now clear that the dominant resonance close to the Sun is, by far, the spiral corotation resonance, and this resonance and its very complex structure (observed in the dynamical map in Fig.~\ref{fig:dynamicalMap}) should be  considered in forthcoming models. In addition, our work is also relevant to the quest for solar siblings, the search for the lost members of the solar family \citep{ramirezEtal2014ApJ}.

Certainly, the interpretation of the kinematics of young stars associated to star-forming regions inside the Local arm (like the Orion complex of molecular clouds and Sco-Cen-Lupus regions, for instance) will have to be re-examined under the light of our new understanding of the  orbits of stars close to corotation. The chemical enrichment barrier separating the regions of the Galaxy situated inside and outside the corotation radius, which gives rise to a metallicity step precisely at corotation, well observed in the sample of open clusters \citep{lepineEtal2011MNRAS}, can now be understood in a new manner. Interestingly, the building up of a metallicity step is also a long-duration process, and in that paper it was used to set  a lower limit to the age of the spiral structure (about 3\,Gyr).

Furthermore, since star formation rate and metallicity enrichment rate are proportional, one might expect the Local arm to be relatively metal rich.
Thus metalliticy variations should not only be observed in the radial direction, but also in the  azimuthal direction. The metallicity should be larger close to the center of the banana-like structure of the island of stability, where star formation occurred, and smaller near the frontiers
between neighboring banana structures. Indeed, this ``azimuthal gradient'' has been observed \citep{lepineEtal2011MNRAS} and can now be better understood.


As a final remark, many authors have discussed the question of habitable zones in the Galaxy  \citep{marochnik1983Ap,balazs1988ASSL,gonzalez2005OLEB,diasLepine2005ApJ}, some of them giving theoretical arguments which locate the Galactic Habitable Zone (GHZ), as well as the solar orbit, close to the corotation circle. In addition, chemical evolution models of the Galaxy support the hypothesis that the GHZ is restricted to a narrow region centered at the Sun's radius \citep{lineweaverEtal2004Sci}. The passage of the Sun through spiral arms  would be a catastrophic event which might affect Earth's biodiversity and cause mass extinctions \citep{balazs1988ASSL, lineweaverEtal2004Sci, filipovicEtal2013SerAJ}. According to our scenario, the situation is even more favorable to life, since the Sun never crossed the main neighboring spiral arms (Sagittarius and Perseus) in the past, at least for a time equal to the lifetime of the present spiral structure, which is possibly of the order of a few Gyrs. Therefore, this no-crossing condition (if it were essential) restricts the possible Galactic regions for emergence of life to the local corotation zone.



\subsection{A scenario for the evolution of the Local arm}\label{sec:evolution}

We suppose that the present spiral structure, with its Sagittarius-Carina and Perseus arms, was possibly triggered by the collision of our Galaxy with a sufficiently massive extragalactic object some 2 -- 3 Gyr ago, according to an accepted mechanism of spiral arms formation \citep{purcellEtalNature2011,elmegreenIAUSymp2012}. As soon as the new spiral structure emerged, the local corotation zone was established, and many stars and gas were trapped in that zone. At first, this did not represent an increase of mass, but only a change in the density distribution of matter. However, while the stars are collisionless, the same is not true for the gas. Gas clouds going back and forth in both radial and azimuthal directions, following the trapped orbits,  collided, stimulating the formation of new trapped stars. This  process is still going on nowadays, since we see  many recently formed stars (open clusters, masers) in this region. Interestingly, the great majority of the stars which were born recently are trapped in the Local arm. Therefore, we can say that we are witnessing the building up of the Local arm, which has reached its present-day intensity by means of a slow process. This may be an indication that the observed spiral structure was present for a long time (a few Gyrs).

What is the source of the gas? In all the models of chemical evolution of the Galactic disk \citep[e.g.][]{laceyFallMNRAS1983,chiappiniEtalApJ1997,matteuciiASPCS2001}, there is a term of infall of gas onto the disk, which explains why the gas has not yet exhausted. Certainly, most of the gas brought by infall on a Galactic range of radius around corotation was responsible for  the continuation of star formation in the Local arm. The energy dissipation processes tended to produce a settlement of gas in the central region of the stability zone.

In order to make use of the investigation of the kinematics of the Local arm by \cite{liuEtalApJL2017}, we must rename what these authors call the trailing and leading sides of the arm.
In our model, these terms lose their sense, since we consider that the region is at corotation, and furthermore, the pitch angle of the Local arm is small (see \citealp{houHan2014AA}).
The trailing side is the inner side (closer to the Galactic center) and the leading side is the outer side. The different azimuthal velocities of the stars located in the inner side and outer side of the arm agree with the loop-like orbits that we predict. Note that there is a major difference between the Local arm and the main arms, from the point of view of the dynamics of the stellar orbits. In the Local arm the stars belonging to the arm perform loops and are trapped, while the stars of the main arms have long orbits around the Galactic center, even when seen in the frame rotating with velocity $\Omega_p$. We expect that, with the release of the GAIA mission data in near future, we will be able to collect samples of stars with highly	 precise positions and space velocities belonging to the Local arm and to the Sagittarius-Carina arm. This will allow us to perform statistical studies of the  differences in the velocity distribution ($V_R$ and $V_\theta$) in the two types of arm, to bring new evidences of the correctness of the basic results of our model and to refine it.



\begin{acknowledgments}
The authors thank Dr. Jorge Mel\'endez for critical reading of this manuscript and helpful suggestions.
This work was supported by the S\~ao Paulo Research Foundation, FAPESP, and the Brazilian National Research Council, CNPq. RSSV and DAB acknowledge the financial support from FAPESP grants 2015/10577-9 and 2016/18886-3, respectively. This work has made use of the facilities
of the Laboratory of Astroinformatics (IAG/USP, NAT/Unicsul), whose purchase was made possible by FAPESP (grant 2009/54006-4) and the INCT-A.
\end{acknowledgments}


\appendix

\section{Spectral analysis method: dynamical power spectra and dynamical maps}\label{appendix}

The qualitative aspects of the Hamiltonian flow are numerically analysed via the Spectral Analysis Method (for details, see \citealt{michtchenkoEtal2002Icar,ferrazmeloMichtchenkoEtal2005LNP}). An application of this method in the context of galactic dynamics is presented in \citet{michtchenkoVieiraBarrosLepine2017AA}, with some discussion about its fundamentals.
This method is used to distinguish between regular and chaotic motions of dynamical systems and is based on the well-known features of power spectra \citep[plot of the amplitude of the Fourier transform of a time series against frequency, see][]{powell1979}. It involves two main steps. The first step is the numerical integration of the equations of motion defined by the full Hamiltonian (\ref{eq:Hamil}). The second step consists of the spectral analysis of the output of the numerical integrations. The time series giving the variation of stellar orbital elements (e.g. the canonical phase-space coordinates) are Fourier-transformed using a standard fast Fourier transform (FFT) algorithm and the main oscillation modes are identified. For more details about these methods, see e.g. \citet{michtchenkoVieiraBarrosLepine2017AA}.

\subsection{Dynamic power spectra}

In order to see how the main oscillation modes evolve when initial conditions vary, we construct a \emph{dynamic power spectrum} plotting the frequencies of the significant peaks of the power spectra as functions of the parameter describing a particular family of solutions.

An example of the dynamic power spectrum is shown on the top panel in Fig.~\ref{fig:powerSpec}. In this case we analyse the oscillations of the radial (red) and azimuthal (black) coordinates of the object \#14 from Table~\ref{tab:masers} and plot their main frequencies as a function of the pattern speed $\Omega_p$. In the domains of regular motion, these frequencies (as well as their harmonics and possible linear combinations between them) evolve continuously when the value of $\Omega_p$ is gradually varied. When Lindblad resonances are approached, the frequency evolution shows a discontinuity characterized by the erratic scatter of values when chaotic layers associated with separatrices are crossed. The smooth evolution of the frequencies is characteristic of regular motion, while
the erratic scattering of the points is characteristic of chaotic motion. Inside the resonance islands, the frequencies split because of the qualitatively distinct dynamics which is intrinsic of the resonance: the passage through the corotation resonance in Fig.~\ref{fig:powerSpec}\,top illustrates this event. The stable domain of the corotation resonance extends from 25 to 29\,km\,s$^{-1}$\,kpc$^{-1}$ of the values of $\Omega_p$. In this work, to perform numerical integrations of the stellar orbits, we choose $\Omega_p=28.5$\,km\,s$^{-1}$\,kpc$^{-1}$.


Another example of a dynamic power spectrum, shown in the bottom panel in Fig.~\ref{fig:powerSpec}, allows us to assess characteristic times of the stellar dynamics in the corotation resonance. The spectrum shows the evolution of periods of the two independent modes of motion, radial and azimuthal, as functions of  the pattern speed $\Omega_p$. The periods which, by our definition, are just inverse of the frequencies, were calculated in close vicinity of the stable corotation libration centers defined by the value of $\Omega_p$ and other parameters from Table~\ref{tab:parameters}. For $\Omega_p=28.5$\,km\,s$^{-1}$\,kpc$^{-1}$, the characteristic radial and azimuthal periods of motion are about 160\,Myr and 1.8\,Gyr, respectively.

\subsection{Dynamical maps on representative planes}

The power spectrum of a time series presents peaks corresponding to the main frequencies of the orbit. Regular orbits are quasi-periodic and have few frequency peaks, given by the two independent frequencies, their harmonics and linear combinations. The amplitude of these peaks, however, drops abruptly when we go to high values. Therefore, their power spectra (plot of the amplitude of the Fourier transform against frequency) present only few significant frequency peaks. On the other hand, chaotic orbits are not confined to an invariant torus; they span a region with higher dimensionality than that of the invariant tori. In practice, this means that their power spectra present a quasi-continuum of frequencies, all of them with comparable magnitudes. Therefore, the number of significant frequencies (defined here as those with amplitude higher than 5\% of the largest peak in the spectrum) is a quantifier of chaos. This number is called \textit{spectral number $N$}; small values of $N$ indicate regular motion, while large values correspond to the onset of chaos. The spectral number $N$ also depends  on the integration time span; the chosen total integration time should be  large enough to  allow the  chaos generated by resonances to be noticeable. It is worth remarking that the method is robust against small variations of the minimum peak amplitude.




\begin{deluxetable}{cccccccc}
\tabletypesize{\scriptsize}
\tablecaption{Orbital elements for the 47 masers associated with the Local arm, and for the Sun.\label{tab:masers}}
\tablewidth{0pt}
\tablenum{1}
\tablecolumns{8}
\tablehead{
\colhead{N$^{\underline{\circ}}$} & \colhead{Source} & \colhead{$R$} & \colhead{$\varphi$} & \colhead{$V_R$} & \colhead{$V_{\theta}$} & \colhead{Emission} & \colhead{Ref.} \\
\colhead{} & \colhead{} & \colhead{(kpc)} & \colhead{($^{\circ}$)} & \colhead{(km s$^{-1}$)} & \colhead{(km s$^{-1}$)} & \colhead{} & \colhead{}
}
\colnumbers
\startdata
 1 & G059.78+00.06 &  7.161 $\pm$  0.024 & 74.89 $\pm$  0.72 &  -1.37 $\pm$  4.32 & 229.02 $\pm$  3.71 & M$^{**}$ & 1 \\
 2 & G069.54--00.97 &  7.503 $\pm$  0.004 & 72.09 $\pm$  0.60 &  -7.28 $\pm$  6.56 & 225.02 $\pm$  5.36 & M$^{*}$ & 1 \\
 3 & G074.03--01.71 &  7.716 $\pm$  0.003 & 78.58 $\pm$  0.32 &  -8.49 $\pm$ 14.89 & 223.74 $\pm$  5.49 & W & 1 \\
 4 & G075.76+00.33 &  7.906 $\pm$  0.053 & 64.52 $\pm$  1.93 &  -0.90 $\pm$ 13.81 & 215.67 $\pm$  9.52 & W & 1 \\
 5 & G075.78+00.34 &  7.976 $\pm$  0.103 & 62.25 $\pm$  3.08 &   3.37 $\pm$ 13.69 & 226.92 $\pm$  6.61 & W & 1 \\
 6 & G076.38--00.61 &  7.797 $\pm$  0.007 & 80.68 $\pm$  0.65 &  -3.02 $\pm$ 25.74 & 219.20 $\pm$  5.70 & W & 1 \\
 7 & G078.12+03.63 &  7.829 $\pm$  0.001 & 78.20 $\pm$  0.59 &  35.92 $\pm$  5.88 & 216.59 $\pm$  5.38 & W, M$^{*}$ & 1 \\
 8 & G078.88+00.70 &  8.052 $\pm$  0.059 & 66.03 $\pm$  1.85 &  12.44 $\pm$ 14.96 & 219.77 $\pm$  7.89 & W & 1 \\
 9 & G079.73+00.99 &  7.872 $\pm$  0.001 & 80.24 $\pm$  0.83 &  -2.13 $\pm$  5.65 & 220.20 $\pm$  5.38 & M$^{*}$ & 1 \\
10 & G079.87+01.17 &  7.878 $\pm$  0.002 & 78.38 $\pm$  0.51 &  -9.40 $\pm$ 13.94 & 218.73 $\pm$ 10.19 & W & 1 \\
11 & G080.79--01.92 &  7.904 $\pm$  0.005 & 78.39 $\pm$  0.89 &   2.19 $\pm$  6.26 & 220.92 $\pm$  3.59 & W, S & 1 \\
12 & G080.86+00.38 &  7.901 $\pm$  0.002 & 79.52 $\pm$  0.58 &  -7.74 $\pm$  4.74 & 221.30 $\pm$  5.38 & M$^{*}$ & 1 \\
13 & G081.75+00.59 &  7.925 $\pm$  0.004 & 79.19 $\pm$  0.57 &  -0.26 $\pm$  4.74 & 221.75 $\pm$  3.60 & M$^{*}$ & 1 \\
14 & G081.87+00.78 &  7.921 $\pm$  0.001 & 80.68 $\pm$  0.51 &   0.36 $\pm$  4.51 & 231.60 $\pm$  3.60 & M$^{*}$ & 1 \\
15 & G090.21+02.32 &  8.031 $\pm$  0.002 & 85.19 $\pm$  0.12 &   3.95 $\pm$  7.37 & 223.87 $\pm$  5.40 & W & 1 \\
16 & G092.67+03.07 &  8.238 $\pm$  0.013 & 78.61 $\pm$  0.36 &  16.16 $\pm$  6.95 & 224.24 $\pm$ 10.01 & W & 1 \\
17 & G105.41+09.87 &  8.275 $\pm$  0.018 & 84.17 $\pm$  0.31 &  -8.44 $\pm$  6.95 & 228.98 $\pm$  5.65 & W & 1 \\
18 & G107.29+05.63 &  8.263 $\pm$  0.024 & 84.88 $\pm$  0.41 &  -5.20 $\pm$  7.00 & 222.86 $\pm$  5.71 & W & 1 \\
19 & G108.18+05.51 &  8.274 $\pm$  0.036 & 84.91 $\pm$  0.58 &  -0.22 $\pm$  5.51 & 222.09 $\pm$  4.01 & M$^{*}$ & 1 \\
20 & G121.29+00.65 &  8.519 $\pm$  0.020 & 84.66 $\pm$  0.18 & -10.40 $\pm$  4.29 & 219.94 $\pm$  4.96 & M$^{*}$ & 1 \\
21 & G176.51+00.20 &  8.962 $\pm$  0.019 & 89.63 $\pm$  0.01 & -15.80 $\pm$  5.09 & 214.51 $\pm$  2.50 & W & 1 \\
22 & G209.00--19.38 &  8.344 $\pm$  0.004 & 91.30 $\pm$  0.02 &   1.36 $\pm$  4.49 & 228.91 $\pm$  3.23 & S & 1 \\
23 & G232.62+00.99 &  9.116 $\pm$  0.071 & 98.41 $\pm$  0.43 &  -2.18 $\pm$  3.13 & 226.34 $\pm$  3.60 & M$^{**}$ & 1 \\
24 & G239.35--05.06 &  8.652 $\pm$  0.047 & 96.65 $\pm$  0.41 &  -0.00 $\pm$  2.40 & 220.41 $\pm$  3.33 & W & 1 \\
25 & G031.56+05.33 &  7.651 $\pm$  0.003 & 88.38 $\pm$  0.01 &  -0.74 $\pm$  2.70 & 232.91 $\pm$  2.59 & C & 2 \\
26 & G071.31+00.83 &  7.872 $\pm$  0.155 & 55.61 $\pm$  4.01 &  -6.79 $\pm$ 13.28 & 235.40 $\pm$  6.58 & W & 2 \\
27 & G071.33+03.07 &  7.612 $\pm$  0.011 & 76.67 $\pm$  0.85 & -19.69 $\pm$  1.77 & 227.27 $\pm$  5.35 & C & 2 \\
28 & G073.12--02.09 &  7.656 $\pm$  0.001 & 72.61 $\pm$  1.03 & -42.04 $\pm$  2.12 & 235.16 $\pm$  3.03 & C & 2 \\
29 & G074.56+00.85 &  7.735 $\pm$  0.021 & 70.15 $\pm$  2.03 &   2.00 $\pm$  5.92 & 218.46 $\pm$  2.25 & W & 2 \\
30 & G158.06--21.42 &  8.201 $\pm$  0.015 & 89.44 $\pm$  0.04 &  11.12 $\pm$  2.95 & 212.62 $\pm$  3.38 & W & 2 \\
31 & G158.35--20.56 &  8.205 $\pm$  0.015 & 89.43 $\pm$  0.04 &  10.63 $\pm$  4.58 & 229.32 $\pm$  2.93 & W & 2 \\
32 & G168.22--16.34 &  8.122 $\pm$  0.003 & 89.82 $\pm$  0.01 &   6.07 $\pm$  1.11 & 230.88 $\pm$  2.03 & C & 2 \\
33 & G168.84--15.52 &  8.126 $\pm$  0.001 & 89.83 $\pm$  0.01 &   4.00 $\pm$  1.89 & 229.21 $\pm$  2.03 & C & 2 \\
34 & G169.37--15.03 &  8.122 $\pm$  0.001 & 89.84 $\pm$  0.01 &   4.90 $\pm$  1.74 & 228.81 $\pm$  2.02 & C & 2 \\
35 & G175.73--16.24 &  8.154 $\pm$  0.001 & 89.92 $\pm$  0.01 &   7.49 $\pm$  1.99 & 227.28 $\pm$  2.01 & C & 2 \\
36 & G176.23--20.89 &  8.137 $\pm$  0.001 & 89.94 $\pm$  0.01 &   8.00 $\pm$  1.50 & 230.64 $\pm$  2.00 & C & 2 \\
37 & G203.32+02.05 &  8.682 $\pm$  0.050 & 91.93 $\pm$  0.13 &  -4.30 $\pm$  4.35 & 217.06 $\pm$  8.18 & W & 2 \\
38 & G208.99--19.38 &  8.347 $\pm$  0.004 & 91.31 $\pm$  0.02 &  -0.75 $\pm$  4.30 & 215.76 $\pm$  2.98 & S & 2 \\
39 & G353.02+16.98 &  7.884 $\pm$  0.005 & 90.10 $\pm$  0.01 &  -1.82 $\pm$  3.02 & 220.52 $\pm$  2.36 & C & 2 \\
40 & G353.10+16.89 &  7.889 $\pm$  0.006 & 90.10 $\pm$  0.01 &  -5.71 $\pm$  3.02 & 227.86 $\pm$  2.25 & C & 2 \\
41 & G353.94+15.84 &  7.829 $\pm$  0.034 & 90.13 $\pm$  0.03 &  -3.83 $\pm$  4.90 & 210.31 $\pm$  6.84 & W & 2 \\
42 & G059.47--00.18 &  7.232 $\pm$  0.025 & 77.14 $\pm$  0.63 & -15.24 $\pm$ 13.00 & 227.94 $\pm$  5.42 & W & 2 \\
43 & G071.52--00.38 &  7.663 $\pm$  0.024 & 63.46 $\pm$  1.25 &   3.87 $\pm$  2.00 & 228.36 $\pm$  3.54 & \nodata & 2 \\
44 & G108.18+05.51 &  8.327 $\pm$  0.011 & 84.08 $\pm$  0.17 &   0.94 $\pm$  1.96 & 223.36 $\pm$  3.41 & M$^{*}$ & 2 \\
45 & G109.87+02.11 &  8.318 $\pm$  0.007 & 84.63 $\pm$  0.11 &  -4.65 $\pm$  1.81 & 226.50 $\pm$  3.38 & M$^{*}$ & 2 \\
46 & G213.70--12.60 &  8.709 $\pm$  0.013 & 93.06 $\pm$  0.05 &   5.48 $\pm$  2.75 & 236.28 $\pm$  2.61 & M$^{*}$ & 2 \\
47 & G213.88--11.84 &  8.739 $\pm$  0.034 & 93.19 $\pm$  0.13 &   2.97 $\pm$  1.38 & 238.27 $\pm$  2.14 & C & 2 \\
\tableline
48 & Sun & 8.0 & 90.0 & -11.1 & 242.24 & \nodata & \nodata \\
\enddata
\tablecomments{Columns are: (1) and (2) - the number and the Galactic source name/coordinates, respectively; (3) and (4) - the Galactic radius and the azimuthal angle in the plane of the Galaxy, respectively; (5) and (6) - the radial velocity and the tangential velocity in the Galactocentric reference frame, respectively; (7) - the type of emission reported in the original papers: ``C'' for continuum at 8.42 GHz, ``M$^{*}$'' for 6.7 GHz methanol maser and ``M$^{**}$'' for 12 GHz methanol maser, ``S'' for 43 GHz SiO maser, ``W'' for 22 GHz H$_2$O maser; and (8) - the references to the papers that published the compilations of maser data: (1) \cite{reidEtal2014ApJ}, (2) \cite{rastorguevEtal2016}. The last row gives the orbital elements of the Sun.}
\end{deluxetable}





\end{document}